# Synthesis of atomically thin hexagonal boron nitride films on nickel foils by molecular beam epitaxy

S. Nakhaie, J.M. Wofford, T. Schumann,[a] U. Jahn, M. Ramsteiner, M. Hanke, J.M.J. Lopes,[b] H. Riechert

*Paul-Drude-Institut für Festkörperelektronik, Hausvogteiplatz 5-7, 10117 Berlin, Germany*

## Abstract

Hexagonal boron nitride (h-BN) is a layered two-dimensional material with properties that make it promising as a dielectric in various applications. We report the growth of h-BN films on Ni foils from elemental B and N using molecular beam epitaxy. The presence of crystalline h-BN over the entire substrate is confirmed by Raman spectroscopy. Atomic force microscopy is used to examine the morphology and continuity of the synthesized films. A scanning electron microscopy study of films obtained using shorter depositions offers insight into the nucleation and growth behavior of h-BN on the Ni substrate. The morphology of h-BN was found to evolve from dendritic, star-shaped islands to larger, smooth triangular ones with increasing growth temperature.

The previous decade has seen extensive research efforts focused on the novel properties of two-dimensional materials and their associated potential applications. This surge in interest was instigated by the isolation of monolayer graphene for the first time,[1] and has rapidly spread to other materials.[2] One such material, hexagonal boron nitride (h-BN), has been the subject of particular attention.

This intense research interest has been driven by the suitability of h-BN for integration into heterostructures with other two-dimensional materials, such as graphene.[3] The first, and perhaps most intuitive way to integrate h-BN and graphene is in vertically stacked heterostructures. In this configuration, the h-BN acts as an insulating layer between electrically active graphene sheets, as well as offering the capability to tune the properties of the graphene layers through moiré effects and other interlayer interactions.[4] This scheme is enhanced by the atomically smooth surface and homogeneous charge potential offered by h-BN, as illustrated by the order of magnitude higher charge carrier mobility in graphene supported by h-BN when compared to $SiO_2$.[3,5,6] Furthermore, h-BN is isomorphic to graphene with a small lattice mismatch of 1.7%, which is commonly regarded as a prerequisite for the growth of defect free epitaxial heterostructures. The structural similarities between the two materials also allow for a second type of heterostructure in which graphene and h-BN are incorporated in a single, laterally patterned monolayer, offering yet more intriguing possibilities.[7,8] Hexagonal boron nitride's large direct band-gap, low dielectric constant, outstanding thermal, mechanical, and chemical stability[9–13] add to its attractiveness for various other research and technological uses.

---

[a] Author's current address: Materials Department, University of California, Santa Barbara, California 93106-5050, USA

[b] Author to whom correspondence should be addressed. Electronic mail: lopes@pdi-berlin.de

Similar to graphene synthesis, optimizing the crystalline quality of h-BN films and establishing control over their thickness while using a scalable synthesis method has proved challenging. Chemical vapor deposition (CVD) is the most thoroughly explored method for the growth of h-BN films. The synthesis of mono- and few-layer h-BN by CVD on catalytic single- or polycrystalline transition metals such as Ni,[14–17] Cu,[18] and Pt[19,20] has been reported. Moreover, using metal organic precursors, synthesis of thicker films of h-BN on dielectrics such as $Al_2O_3$ and 6H-SiC has been achieved.[21,22] Different forms of physical vapor deposition (PVD), such as pulsed laser deposition,[23] and reactive magnetron sputtering,[24] have also been used for h-BN thin film growth. While these studies exhibit the progress being made in h-BN growth, a single scalable synthesis method which combines high-crystalline quality with absolute thickness control remains elusive.

Among the various approaches to synthesizing h-BN films, molecular beam epitaxy (MBE) offers precise control over growth conditions. Whereas CVD growth relies on the decomposition of a molecular precursor, the rate of which may change dramatically with the exposed surface area of the catalytic substrate,[25] MBE faces no such constraints. The absence of catalytic processes in MBE growth is crucial for vertically stacked heterostructures, where the non-catalytic surfaces of graphene, h-BN or other dielectrics will serve as substrates for growth. The precise fluxes and resulting growth rates also offered by MBE facilitate the deposition of sub-monolayer films, making the production of laterally patterned monolayer heterostructures feasible. Similar considerations make MBE ideal for fundamental studies of h-BN crystal growth, the results from which may also benefit synthesis by other methods.

However, despite the sub-monolayer precision it allows, the few existing reports of boron nitride growth by MBE focus on the cubic phase or thicker h-BN films for optoelectronic purposes.[26,27] These experiments have only peripheral bearing on the more precise type of growth pursued here. The present letter reports the synthesis of atomically thin films of h-BN on polycrystalline Ni foils via MBE from elemental precursors. Growth parameters which yield continuous h-BN films are included. Finally, we examine films where deposition was halted prior to h-BN covering the entire Ni foil to offer brief comments regarding the h-BN growth mode on this surface.

All samples were grown in an MBE system with a base pressure of $1.0 \times 10^{-10}$ mbar. Polycrystalline Ni foils (Alpha Aesar, 99.994% pure, 100 μm thick) cut into $1 \times 1$ cm$^2$ pieces were used as substrates. Prior to h-BN deposition, the foils were cleaned using conventional solvents and water, followed by cyclic annealing in UHV and sputtering with Ar (anneal for 30 minutes at 1000 °C, sputter for 20 minutes with a 2 kV acceleration voltage and $10^{-4}$ mbar Ar pressure at 600 °C). Films were synthesized using substrate temperatures ranging from 730 °C to 835 °C, as measured by an optical pyrometer. Growth times ranged from 3 to 5 hours, which allowed for the modulation of the h-BN surface coverage (from isolated islands to full surface coverage). A high-temperature effusion cell (TUBO cell, CreaTec Fischer & Co. GmbH) operating at 1850 °C was used to provide the beam of elemental B. Active N-species were generated using an RF plasma source operating at 350 W with 0.2 sccm of $N_2$ flow, which resulted in a chamber pressure of $1.1 \times 10^{-5}$ mbar during growth. The resulting h-BN films were then characterized using Raman spectroscopy, atomic force microscopy (AFM), and scanning electron microscopy (SEM). Synchrotron-based x-ray reflectivity (XRR) was used to extract the average film thickness.

The growth of crystalline h-BN on the Ni substrates was first confirmed using Raman spectroscopy. Figure 1 depicts representative Raman spectra of h-BN samples synthesized over 3 hours (at 730 and 835 °C) and 5 hours (at 730 °C). The Raman spectra are superimposed to a relatively strong background originating from the Ni substrate. For the spectra shown in Figure 1, this background signal has been subtracted (a spectrum before background subtraction is illustrated in the inset of Figure 1). Figure 1 also depicts a spectrum from commercially available h-BN powder for comparison. Note that for the cases of 3 hours synthesis, isolated h-BN islands grow on the Ni surface (see Figure 4 and detailed discussions later), while for the 5 hour long growth a closed film is formed. The latter is confirmed by the uninterrupted observation of the h-BN Raman signal over the entire substrate surface. Regardless of the surface coverage, the sharp peak observed in all spectra, which arises from the doubly degenerate in-plane optical phonons of h-BN with $E_{2g}$ symmetry, is indicative of high-quality h-BN.[28] This peak showed slight width variations in numerous Raman measurements, ranging from 10 cm$^{-1}$ to 18 cm$^{-1}$. The intensity of the peak exhibits some variations, likely as a result of an inhomogeneous thickness distribution. The average peak height to background ratio for the continuous films is 0.18, while for the ramified and triangular islands this value is 1.1 and 3.7, respectively. Since the intensity of the background originating from the underlying Ni is always on the same order of magnitude, the variation in the relative intensity of the characteristic h-BN peak to the background reflects a change in the volume of h-BN being probed. Finally, we also observed small shifts in the peak position for the different types of h-BN structures. Variation in strain is a potential explanation for the observed shifts. [29]

The surface morphology resulting from a 5 hour synthesis at 730 °C (continuous film) was evaluated by AFM. The root-mean-square (RMS) roughness of the h-BN film was measured to be 1.0 and 0.3 nm over $1 \times 1$ μm$^2$ and $0.1 \times 0.1$ μm$^2$ areas, respectively. Uneven regions of the underlying Ni surface resulted in higher RMS values in some areas. Profile measurements from edges of the h-BN film showed 0.5-1.5 nm steps down to the bare Ni, revealing that the film is composed of few atomic layers of h-BN (the interlayer distance in h-BN is ~ 0.3 nm).[30] AFM also allowed the continuity of the h-BN films to be confirmed, as we now describe. In addition to the topographic features of the underlying Ni foil, such as step edges, terraces, etc., a cellular array of linear features is easily discernible which are identified as wrinkles (Figure 2, red arrows). The h-BN film is expected to develop wrinkles during sample cooling due the relaxation of compressive strain in the h-BN arising from the unequal expansion coefficients of h-BN[31] and Ni.[32] Such wrinkles have been observed previously in both h-BN and graphene grown on metal substrates.[18,20,33] These wrinkles were not present on the bare Ni surface. The ubiquity of the wrinkle structure in numerous AFM scans, together with the uninterrupted observation of the h-BN Raman signal, offer strong evidence of a continuous h-BN film.

AFM scans also show a second distinct type of feature in the h-BN films. Rather than having a smooth profile along their length as the h-BN wrinkles do, these rougher raised regions appear to be decorated with a line of discrete "dots". The "dotted" features also protrude higher out of the plane of the film than the wrinkles (typically ~15 nm versus ~5 nm - see inset in Figure 2a), and are sufficiently pronounced to be evident even in larger area AFM scans (Figure 2b). These distinctions are sufficient to suggest that the two types of features form due to different mechanisms. Although this alternative mechanism is not yet clear, we speculate the dotted ridges may mark grain boundaries in a polycrystalline h-BN film. For instance, surface contaminants or excess surface B (e.g. due to B-rich growth conditions) could be pushed by the growth front of the expanding h-BN crystal, with the consequent accumulation of material at grain boundaries leading to the formation of the observed "dots".

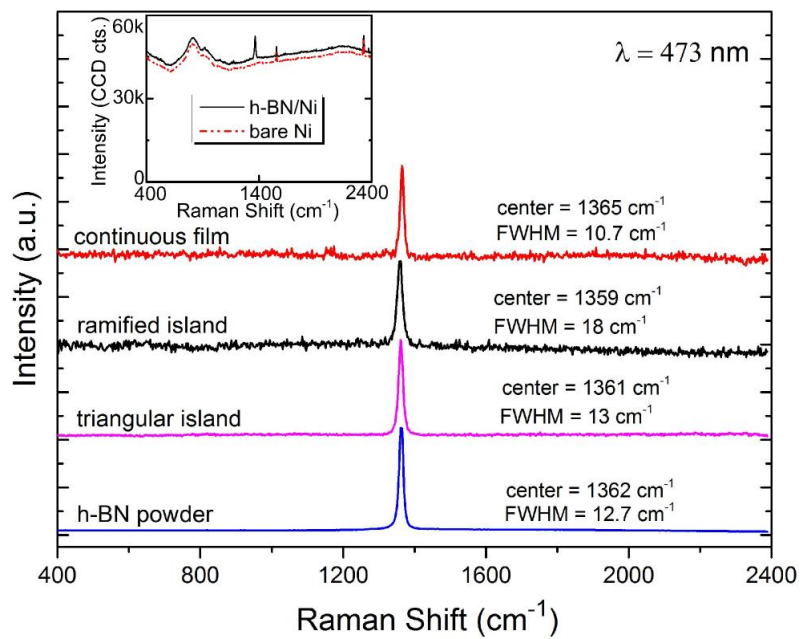

Figure 1. Raman spectra collected from a continuous h-BN film on Ni synthesized at 730 °C over 5 hours (red), and from h-BN islands grown for 3 hours at 730 °C (black) and 835 °C (magenta). The substrate-related background signal was substracted. The inset shows the Raman spectrum from the continuous film before background subtraction as well as the spectrum from a bare Ni substrate (an offset was applied for better visualization). A Raman spectrum from h-BN powder (blue) is plotted for comparison. The spectra were excited at a wavelength of 473 nm with the laser beam focused onto the sample surface to a ~ 1 um diameter spot.

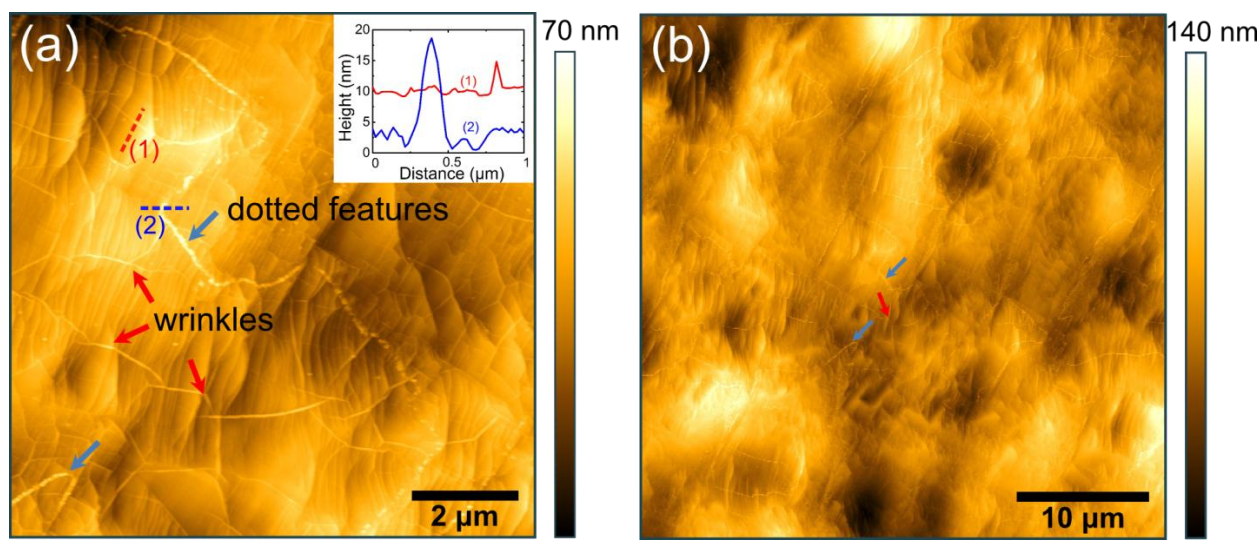

Figure 2. AFM images of a continuous h-BN film grown on Ni (730 °C, 5 hours). (a) Two distinct types of ridges, marked with red and blue arrows, are observed across the entire area of the film. The smooth, continuous ridges are wrinkles in h-BN (red arrows), while the objects marked with blue arrows are discontinuous and appear dotted. The inset shows the line scans (1) and (2) obtained across a wrinkle and a dotted feature, respectively. The dotted features are ~15 nm high, while the wrinkles ~ 5 nm. (b) Larger area AFM scan of the same sample region. The blue and red arrows indicate the same features as in (a).

Synchrotron-based XRR has been employed to estimate the thickness of the continuous h-BN films. Since the layer is anticipated to be extremely (i.e., atomically) thin - in conjunction with its low-Z, and thus weakly scattering constituents (boron and nitrogen) - highly brilliant radiation is mandatory for the successful application of this technique. Figure 3a shows the reflectivity curve as measured at the U125/2-KMC beam-line (BESSYII, Helmholtz-Zentrum Berlin). An x-ray energy of 10 keV has been selected by a double Si(111) monochromator, which yields an energy resolution ($\Delta E/E$) of about $10^{-4}$. Due to the roughness of the underlying Ni foil, as well as the out-of-plane wrinkles and dotted features, the reflected intensity rapidly decreases with increasing the angle of incidence, $\alpha_i$. Nevertheless, thickness oscillations are clearly traceable in the experiment. Figure 3b depicts three kinematic scattering simulations for h-BN layer thicknesses between 0.8 and 1.0 nm on a Ni substrate. From a direct comparison we can conclude that the averaged h-BN layer thickness is close to 0.9 nm. This shows that the h-BN film is about three atomic layers thick (a single layer corresponds to ~ 0.3 nm), in agreement with the AFM profile measurements taken at edges of the h-BN film.

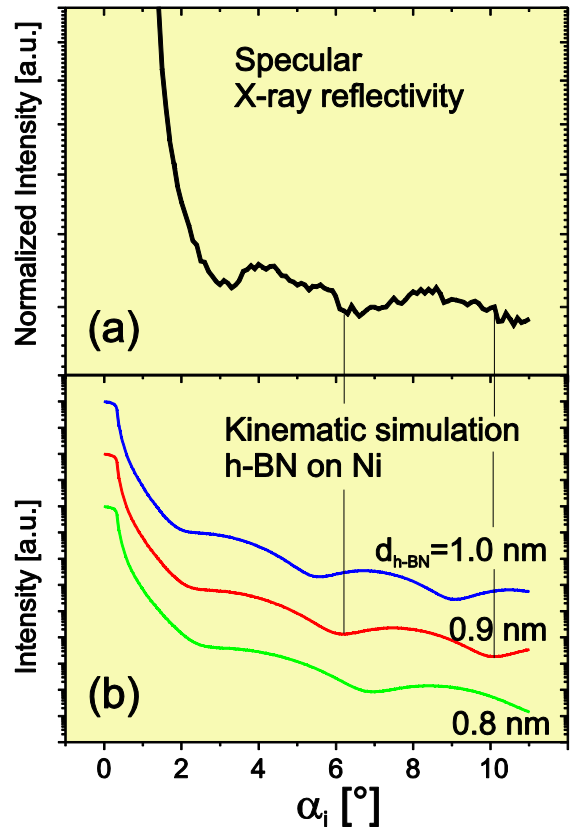

Figure 3. (a) Experimentally observed specularly reflected x-ray intensity collected from a continuous h-BN film on Ni synthesized at 730 °C over 5 hours. (b) Kinematic scattering simulations for a h-BN layer with the thickness ($d_{h-BN}$) varying between 0.8 and 1.0 nm on a Ni substrate.

Using shorter duration depositions we were able to gain insight into the nucleation and growth behavior of h-BN prior to the formation of a closed film. Growth times of 3 hours or shorter produced discrete regions of h-BN which were easily identifiable by SEM (Figure 4). While the atomistic growth mechanism of h-BN on Ni is unknown, a close examination of the island structures offers some clues to their origin. For instance, the h-BN islands typically include a small region of prominent contrast at the approximate geometric center of the island, possibly marking a surface defect at which the island nucleated heterogeneously. Heterogeneous nucleation has also been observed previously during the growth of other two-dimensional materials, such as graphene.[34] The surface imperfections at which nucleation occurs here may be clusters of excess surface B, or possibly isolated regions of the intermetallic $Ni_3B$ phase.[35,36]

The morphology of the h-BN islands changes dramatically with the growth temperature (Figure 4). Hexagonal boron nitride islands grown at a substrate temperature of 730 °C have a distinct "star"-shape, with multiple elongated lobes extending from a common nucleation site over few-micron lengths. With the same growth duration but a 835 °C substrate temperature, the h-BN instead forms smooth, compact, approximately triangular islands which are almost an order of magnitude larger. However, isolated regions with shapes other than those depicted in Figure 4 have also been observed, possibly due to changes in h-BN growth behavior on different crystallographic orientations of Ni.[15]

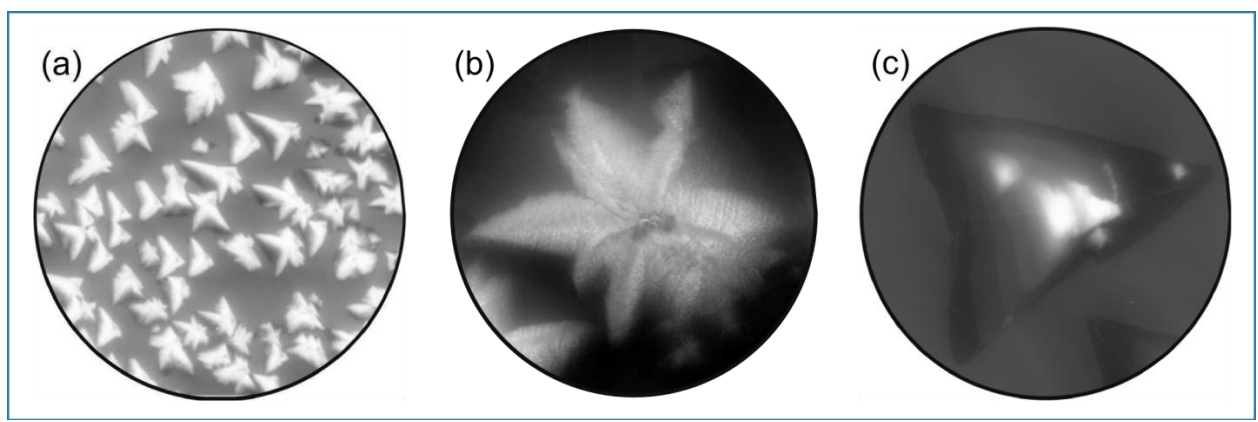

Figure 4. SEM micrographs of h-BN grown on a Ni foil by MBE. In these instances growth was halted prior to the formation of a closed film. A substrate temperature of 730 °C results in star-shaped h-BN islands (bright contrast in (a) and (b)). When the substrate temperature is raised to 835 °C for growth, much larger islands form, which have a smoother, more compact shape (c). The particular large island shown in (c) is also more than a single h-BN layer thick, as the contrast fluctuations show. The field of view in (a) is 30 μm, 9 μm in (b), and 70 μm in (c).

It is interesting to note the superficial similarities between the star-shaped h-BN islands formed on Ni and the morphology of graphene growing on (001) oriented grains in Cu foils. The fairly consistent orientation of the long-axes of the h-BN island lobes is a particularly striking detail to have in common with graphene islands on Cu. In the case of graphene on Cu, the island shape results from a complex interplay between the attachment limited growth mode, the polycrystalline island structure, and the substrate surface crystallography.[34] Whether the h-BN island shape observed here is caused by analogous growth phenomena remains an open question. The star-shaped islands could, for instance, be the product of dendritic growth where the adatom diffusion length is of the same order of magnitude as the island size. Indeed, the dramatic transformation in island morphology with growth temperature is more consistent with this possible explanation.

The morphology of the larger, compact h-BN structures grown at higher substrate temperatures also has precedent in the literature. Triangular h-BN islands have been reported for CVD grown h-BN on both Ni[16,17] and Cu[18] substrates, and theoretical calculations suggest this is due to the lower free energy of the N-terminated crystal edge.[37] However, because of their significantly larger size, multi-layer thickness, random orientation, and deviation from perfectly triangular shapes, it is unlikely that the preference for a given edge termination fully explains the morphology presented in Figure 4c. Rather the dramatic change in island morphology likely results from an interplay between kinetic factors, such as increased adatom diffusion, in combination with the energetic anisotropy of the edge structure. Contrast variations in the interior of the large island in Figure 4c reveal its layered structure. As the secondary electron intensity emitted from h-BN scales linearly with the number of layers,[38] each observable variation likely marks the perimeter of an additional layer. The multi-layer thickness imaged by SEM corroborates the Raman analysis (see Figure 1), in which the highest peak height to background ratio (~ 3.7) was obtained for this type of island.

In summary, we report the growth of atomically thin h-BN films on Ni foils by MBE using elemental B and N. In these experiments, films were synthesized at substrate temperatures ranging from 730 °C to 835 °C over 3 to 5 hours. The presence of crystalline h-BN was confirmed by Raman spectroscopy, which revealed a sharp peak at ~1365 cm$^{-1}$. AFM was used to examine the surface morphology of the grown films, and revealed two distinct features which were omnipresent over the entire film: a cellular array of linear features identified as wrinkles, and a second rougher type of ridge. The rougher, "dotted" ridges appear to originate from a different, as yet unknown mechanism. Using shorter growth times we were able to gain insight into the nucleation and growth behavior of h-BN before forming a closed film.  According to SEM imaging, the morphology of sub-monolayer islands evolves from ramified and "star"-shaped to much larger, smooth and triangular islands with increasing growth temperature. Scanning electron micrographs also clearly show a small region of prominent contrast in the center of the grown islands, suggesting that the h-BN may have nucleated heterogeneously. Further investigations aiming at the elucidation of the nucleation and growth mechanisms are in progress. The present results serve to demonstrate the potential of MBE as a technique for realizing the controlled growth of continuous h-BN layers over large areas.

## Acknowledgments

The authors would like to thank C Chèze for critical reading of the manuscript; H-P Schönherr, M Höricke, and C Herrmann for technical support; G. Lupina, G. Lippert, and J. Dabrowski from IHP Microelectronics for fruitful discussions and support with Raman measurements; B. Jenichen for help with XRR measurements. JMW acknowledges support from the Alexander von Humboldt Foundation. The authors also acknowledge the Helmholtz-Zentrum Berlin for providing beamtime at the PHARAO Endstation.

# References

[1] K.S. Novoselov, A.K. Geim, S. V Morozov, D. Jiang, Y. Zhang, S. V Dubonos, I. V Grigorieva, and A.A. Firsov, Science **306**, 666 (2004).

[2] K.S. Novoselov, D. Jiang, F. Schedin, T.J. Booth, V. V Khotkevich, S. V Morozov, and A.K. Geim, Proc. Natl. Acad. Sci. U. S. A. **102**, 10451 (2005).

[3] C.R. Dean, A.F. Young, I. Meric, C. Lee, L. Wang, S. Sorgenfrei, K. Watanabe, T. Taniguchi, P. Kim, K.L. Shepard, and J. Hone, Nat. Nanotechnol. **5**, 722 (2010).

[4] M. Yankowitz, J. Xue, D. Cormode, J.D. Sanchez-Yamagishi, K. Watanabe, T. Taniguchi, P. Jarillo-Herrero, P. Jacquod, and B.J. LeRoy, Nat. Phys. **8**, 382 (2012).

[5] W. Gannett, W. Regan, K. Watanabe, T. Taniguchi, M.F. Crommie, and A. Zettl, Appl. Phys. Lett. **98**, 242105 (2011).

[6] E. Kim, T. Yu, E. Sang Song, and B. Yu, Appl. Phys. Lett. **98**, 262103 (2011).

[7] P. Sutter, R. Cortes, J. Lahiri, and E. Sutter, Nano Lett. **12**, 4869 (2012).

[8] L. Liu, J. Park, D.A. Siegel, K.F. McCarty, K.W. Clark, W. Deng, L. Basile, J.C. Idrobo, A.-P. Li, and G. Gu, Science **343**, 163 (2014).

[9] Z. Liu, Y. Gong, W. Zhou, L. Ma, J. Yu, J.C. Idrobo, J. Jung, A.H. MacDonald, R. Vajtai, J. Lou, and P.M. Ajayan, Nat. Commun. **4**, 2541 (2013).

[10] K. Watanabe, T. Taniguchi, and H. Kanda, Nat. Mater. **3**, 404 (2004).

[11] K.K. Kim, A. Hsu, X. Jia, S.M. Kim, Y. Shi, M. Dresselhaus, T. Palacios, and J. Kong, ACS Nano **6**, 8583 (2012).

[12] T. Sugino and T. Tai, Jpn. J. Appl. Phys. **39**, L1101 (2000).

[13] D. Golberg, Y. Bando, Y. Huang, T. Terao, M. Mitome, and C. Tang, ACS Nano **4**, 2979 (2010).

[14] A. Nagashima, N. Tejima, Y. Gamou, T. Kawai, and C. Oshima, Phys. Rev. B **51**, 4606 (1995).

[15] Y.-H. Lee, K.-K. Liu, A.-Y. Lu, C.-Y. Wu, C.-T. Lin, W. Zhang, C.-Y. Su, C.-L. Hsu, T.-W. Lin, K.-H. Wei, Y. Shi, and L.-J. Li, RSC Adv. **2**, 111 (2012).

[16] W. Auwärter, M. Muntwiler, J. Osterwalder, and T. Greber, Surf. Sci. **545**, L735 (2003).

[17] W. Auwärter, H.U. Suter, H. Sachdev, and T. Greber, Chem. Mater. **16**, 343 (2004).

[18] K.K. Kim, A. Hsu, X. Jia, S.M. Kim, Y. Shi, M. Hofmann, D. Nezich, J.F. Rodriguez-Nieva, M. Dresselhaus, T. Palacios, and J. Kong, Nano Lett. **12**, 161 (2012).

[19] A. Nagashima, N. Tejima, Y. Gamou, T. Kawai, and C. Oshima, Surf. Sci. **357-358**, 307 (1996).

[20] J.-H. Park, J.C. Park, S.J. Yun, H. Kim, D.H. Luong, S.M. Kim, S.H. Choi, W. Yang, J. Kong, K.K. Kim, and Y.H. Lee, ACS Nano **8**, 8520 (2014).

[21] R. Dahal, J. Li, S. Majety, B.N. Pantha, X.K. Cao, J.Y. Lin, and H.X. Jiang, Appl. Phys. Lett. **98**, 211110 (2011).

[22] S. Majety, J. Li, W.P. Zhao, B. Huang, S.H. Wei, J.Y. Lin, and H.X. Jiang, Appl. Phys. Lett. **102**, 213505 (2013).

[23] N.R. Glavin, M.L. Jespersen, M.H. Check, J. Hu, A.M. Hilton, T.S. Fisher, and A.A. Voevodin, Thin Solid Films **572**, 245 (2014).

[24] P. Sutter, J. Lahiri, P. Zahl, B. Wang, and E. Sutter, Nano Lett. **13**, 276 (2013).

[25] P. Sutter, J. Lahiri, P. Albrecht, and E. Sutter, ACS Nano **5**, 7303 (2011).

[26] M.J. Paisley, J. Vac. Sci. Technol. B Microelectron. Nanom. Struct. **8**, 323 (1990).

[27] C.L. Tsai, Y. Kobayashi, T. Akasaka, and M. Kasu, J. Cryst. Growth **311**, 3054 (2009).

[28] S. Reich, A. Ferrari, R. Arenal, A. Loiseau, I. Bello, and J. Robertson, Phys. Rev. B **71**, 205201 (2005).

[29] R. V Gorbachev, I. Riaz, R.R. Nair, R. Jalil, L. Britnell, B.D. Belle, E.W. Hill, K.S. Novoselov, K. Watanabe, T. Taniguchi, A.K. Geim, and P. Blake, Small **7**, 465 (2011).

[30] A. Pakdel, Y. Bando, and D. Golberg, Chem. Soc. Rev. **43**, 934 (2014).

[31] R.S. Pease, Acta Crystallogr. **5**, 356 (1952).

[32] T.G. Kollie, Phys. Rev. B **16**, 4872 (1977).

[33] X. Li, W. Cai, J. An, S. Kim, J. Nah, D. Yang, R. Piner, A. Velamakanni, I. Jung, E. Tutuc, S.K. Banerjee, L. Colombo, and R.S. Ruoff, Science **324**, 1312 (2009).

[34] J.M. Wofford, S. Nie, K.F. McCarty, N.C. Bartelt, and O.D. Dubon, Nano Lett. **10**, 4890 (2010).

[35] A.N. Campbell, A.W. Mullendore, C.R. Hills, and J.B. Vandersande, J. Mater. Sci. **23**, 4049 (1988).

[36] K.I. Portnoi, V.M. Romashov, V.M. Chubarov, M.K. Levinskaya, and S.E. Salibekov, Sov. Powder Metall. Met. Ceram. **6**, 99 (1967).

[37] Y. Liu, S. Bhowmick, and B.I. Yakobson, Nano Lett. **11**, 3113 (2011).

[38] P. Sutter and E. Sutter, APL Mater. **2**, 092502 (2014).